\documentclass[conference]{IEEEtran}
\IEEEoverridecommandlockouts
\usepackage{cite}
\usepackage{amsmath,amssymb,amsfonts}
\usepackage{algorithmic}
\usepackage{graphicx}
\usepackage{textcomp}
\usepackage{xcolor}
\usepackage{subfigure}
\usepackage[ruled]{algorithm2e}
\usepackage{multirow}
\usepackage{cases}
\usepackage{booktabs}
\def\BibTeX{{\rm B\kern-.05em{\sc i\kern-.025em b}\kern-.08em
    T\kern-.1667em\lower.7ex\hbox{E}\kern-.125emX}}
\begin{document}

\title{Time-Sensitive Networking over 5G: Experimental Evaluation of a Hybrid 5G and TSN System with \\IEEE 802.1Qbv Traffic}


\author{
\IEEEauthorblockN{Adnan Aijaz and Sajida Gufran}
\IEEEauthorblockA{
\text{Bristol Research and Innovation Laboratory, Toshiba Europe Ltd., Bristol, United Kingdom}\\
firstname.lastname@toshiba-bril.com}
}    

\maketitle

\begin{abstract}
Underpinned by the IEEE 802.1 standards, Time-sensitive networking (TSN)  empowers standard Ethernet to handle stringent real-time requirements of industrial networking. TSN and private 5G will co-exist in industrial systems; hence, converged operation of the two is crucial to achieving end-to-end deterministic performance. This work conducts a testbed-based evaluation of a hybrid 5G and TSN system with over-the-air transmission of scheduled real-time TSN traffic (based on IEEE 802.1Qbv standard). The main objective is to bring the dynamics of hybrid 5G and TSN deployments to spotlight. The testbed comprises off-the-shelf TSN and 5G devices and a near product-grade 5G system. The results show the impact of 802.1Qbv parameters and 5G system capabilities on end-to-end deterministic performance.  
The findings of this study have significance for design and optimization of 3GPP-defined bridge model (black box model) for 5G/TSN integration. 
\end{abstract}


\section{Introduction}
Time-sensitive Networking (TSN) is a set of standards under the IEEE 802.1 working group \cite{tsn} that improve  real-time capabilities of the standard Ethernet \cite{TSN_standards}. TSN provides guaranteed data delivery with deterministic and bounded latency and extremely low data loss. TSN supports both time-critical and best-effort traffic over a single standard Ethernet network. It is expected to be the de-facto wired technology for industrial networks, replacing proprietary technologies. According to a recent report\footnote{https://www.marketsandmarkets.com/Market-Reports/time-sensitive-networking-market-215000493.html}, the global TSN market is expected to reach 1.7 billion USD by 2028. The increased demand for automation in the industrial sector is the key driving factor for TSN market growth. 

However, TSN will co-exist with wireless technologies which are becoming increasingly important for industrial networking \cite{Ind_comm_future}. The fifth-generation
(5G) mobile/cellular technology is capable of handling diverse applications; therefore, it underpins a unified wireless interface for industrial communication \cite{pvt_5G}. In particular, the ultra-reliable low-latency communication (uRLLC) capability of 5G is promising for time-sensitive traffic. 

Converged operation of 5G and TSN systems is crucial to end-to-end deterministic connectivity and performance guarantees over hybrid wired and wireless domains.  3GPP has defined the bridge model for integration of 5G and TSN, wherein the 5G system appears as a black box to TSN entities \cite{3gpp_23_734}. The 5G system handles the service requirements of TSN traffic through its internal protocols. To achieve TSN-like functionality, a number of challenges need to be addressed \cite{pvt_5G}, including resource management techniques for handling TSN traffic.


\subsection{Related Work}
Nasrallah \emph{et al.} \cite{TSN_survey} carried out a comprehensive survey of TSN standards and ultra-low latency mechanisms in 5G. Most existing studies targeting converged operation of 5G and TSN systems rely on simulation techniques. Martenvormfelde \emph{et al.} \cite{RW_sim}  and Debnath \emph{et al.} \cite{RW_sim2} have developed  simulation-based frameworks for 5G and TSN integration. Rost and Kolding \cite{RW_sim3} conducted simulations-based latency evaluation of an integrated 5G and TSN system. Testbed-based evaluation for convergence of 5G and TSN is still in infancy. Senk \emph{et al.} survey open-source projects for potential 5G and TSN integration. Guiraud \emph{et al.} developed a TSN switch prototype.  Seijo \emph{et al.} developed a hybrid wired/wireless TSN device prototype with non-3GPP wireless technologies.  Ulbricht \emph{et al.} \cite{RW_tb4} introduced the TSN-FlexTest testbed for TSN-related experimentation. 

Various studies (e.g., \cite{RW_sch1, RW_sch2}) have investigated the problem of radio resource allocation techniques for 5G and TSN integration. 
Analysis and enhancements for time synchronization in integrated 5G and TSN systems has been addressed in different studies as well \cite{Eval_time_synch, RW_synch2, RW_synch3}.


\subsection{Contributions and Outline}
To this end, this study adopts an experimental approach toward integrated 5G and TSN systems and conducts a testbed-based evaluation. The primary objective is to understand the dynamics of hybrid wired and wireless networking with TSN traffic sent via a 5G system.  The key contributions are summarized as follows. 

\begin{itemize}
\item We create a hybrid TSN and 5G testbed based on off-the-shelf TSN and 5G devices and a near product-grade end-to-end 5G system\footnote{We consider a vanilla 5G system with no specific enhancements or optimizations for providing TSN-like functionality.}. 

\item We generate real TSN traffic, based on IEEE 802.1Qbv specifications \cite{802_1Qbv}, which is transmitted over-the-air to a 5G system. 

\item We conduct performance evaluation with an emphasis on end-to-end deterministic behavior under different parameters of scheduled 802.1Qbv traffic.  
\end{itemize}

We adopt IEEE 802.1Qbv as it is the most widely used TSN standard for industrial networking. It also creates the most stringent performance requirements due to cyclic traffic patterns, bounded low latency, and zero jitter. 
Our experimental evaluation provides insights for system-level enhancements and end-to-end resource allocation and optimization of a 5G system in order to effectively handle scheduled TSN traffic. 


The rest of the paper is organized as follow. Section \ref{sect_prelim} captures some prerequisites on 802.1Qbv. The testbed setup is described in Section \ref{sect_tb}. Experimental evaluation and results are presented in Section \ref{sect_exp}. Section \ref{sect_insights} highlights some key insights and takeaways from our experimentation. Finally, the paper is concluded in Section \ref{sect_cr}. 

\section{Overview of the IEEE 802.1Qbv Standard} \label{sect_prelim}
The set of IEEE TSN standards can be broadly classified into: (i) time synchronization (e.g., 802.1AS for timing and synchronization), (ii) ultra-high reliability (802.1CB for frame replication and elimination), (iii) resource management (e.g., 802.1Qcc for TSN configuration), and (iv) bounded low latency (e.g., 802.1Qbv for scheduled traffic).  A review of these TSN standards is beyond the scope of this paper. The interested readers are referred to \cite{TSN_standards} and \cite{TSN_survey} for further details.

The IEEE 802.1Qbv standard \cite{802_1Qbv} specifies a time-aware shaper (TAS) which provides time-aware traffic scheduling with bounded low latency. TAS can perfectly isolate priority classes in time with periodically scheduled traffic gates, thereby enabling ultra low latency applications to coexist with lower priority traffic on the same network.  TAS turns the priority queues into a TDMA-like system through traffic gates governed by a global clock. Traffic is carried in time-triggered windows such that a gate control list (GCL) dictates open or close status for each queue. Guard bands are used with transmit windows to prevent low-priority traffic interfering with high-priority traffic.  Overall TAS requires precise clock synchronization across the entire network and a central network controller to calculate the timings for the traffic gates, based on known propagation delays on the network links. Several cheaper alternatives have been developed as well for asynchronous TAS that needs no clock synchronization, network planning, or a central controller.

\section{Testbed Setup} \label{sect_tb}
\subsection{TSN Gateway}
\begin{figure}
\centering
\includegraphics[scale=0.33]{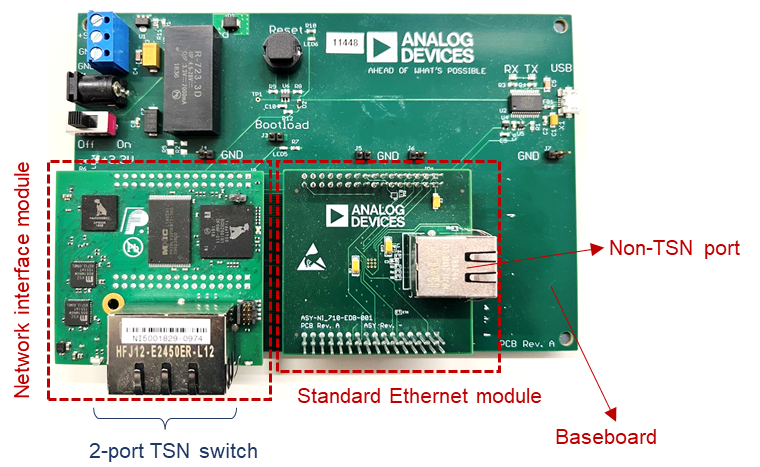} 
\caption{The TSN gateway in our testbed. }
\vspace{-1.5em}
\label{TSN}
\end{figure}

We use the Analog Devices RapID Platform (RAPID-TSNEK-V0001) as a TSN gateway. The TSN gateway (shown in Fig. \ref{TSN}) allows any non-TSN device to participate in a TSN system without native implementation of TSN-specific features. It is based on fido500  real-time Ethernet multi-protocol (REM) programmable 3-port switch
chip that manages TSN functionality. The software on the TSN gateway supports features from different IEEE specifications including 802.1AS (time synchronization), 802.Qbv (scheduled traffic), 802.1Qcc (stream reservation), and stream translation. The 802.1Qcc functionality is implemented through a web server.  

The TSN gateway includes a network interface module which provides a 2-port TSN switch and a standard Ethernet module which provides a non-TSN port. The gateway functionality is provided between a standard 100Base-TX Ethernet port and two 100Base-TX TSN ports. The two TSN ports enable evaluation of any network topology including star, ring, redundant, and line. The TSN gateway is designed for rapid configuration of TSN features, thereby allowing non-TSN applications to exchange TSN data streams with TSN talkers or listeners.  

\subsection{5G System}
\begin{figure}[h]
\centering
\includegraphics[scale=0.35]{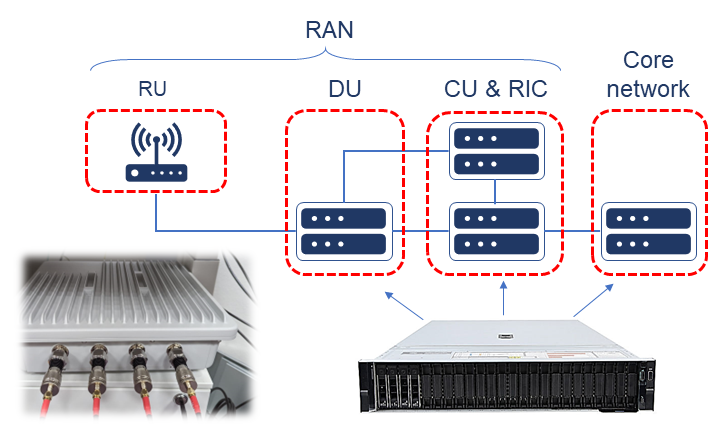} 
\caption{Illustration of the end-to-end 5G system. }
\label{5G_syst}
\vspace{-0.5em}
\end{figure}

The 5G system in our testbed, illustrated in Fig. \ref{5G_syst}, is an integrated multi-vendor 5G standalone system based on O-RAN architecture \cite{o2021ran} and compliant with 3GPP Release 15. The 5G radio access network (RAN) is disaggregated; the gNB is split into the radio unit (RU), the distributed unit (DU), and centralized unit (unit) components. The end-to-end 5G system  also includes core network and the RAN intelligent controller (RIC). Further, the software stack of all components is running on general-purpose high-specification servers with the exception of the RU which is based on proprietary hardware. The RAN and core network segments can be configured via dashboard. 

The 5G system supports time-division duplexing (TDD) and it operates in the band n77u (3.8 - 4.2 GHz), which is the private 5G band in the UK. In terms of performance, the 5G system provides average round-trip latency of around 10 msec, meeting the requirements of most real-time control applications. The supported bandwidths are 40 MHz and 100 MHz, providing peak downlink throughputs of approximately 240 Mbps and  500 Mbps, respectively. The air-interface can be customized for downlink-centric or uplink-centric traffic patterns.  
The 5G systems is field-tested. Further details about the 5G system, including it's performance and field trial, are available in our recent work \cite{BEACON-5G_CSCN}.

\subsection{5G Device}
The primary 5G device in our testbed is a dongle based on a  Quectel RM500Q-GL module, an M.2 to USB adapter, and antennas connected by SMA cables. Compliant with 3GPP Release 15, it supports 5G standalone mode in different sub-6 GHz frequency bands including band n77. The 5G device is provisioned for data connectivity with the 5G system.

\subsection{Configuration of the TSN Gateway}
\begin{figure}
\centering
\includegraphics[scale=0.4]{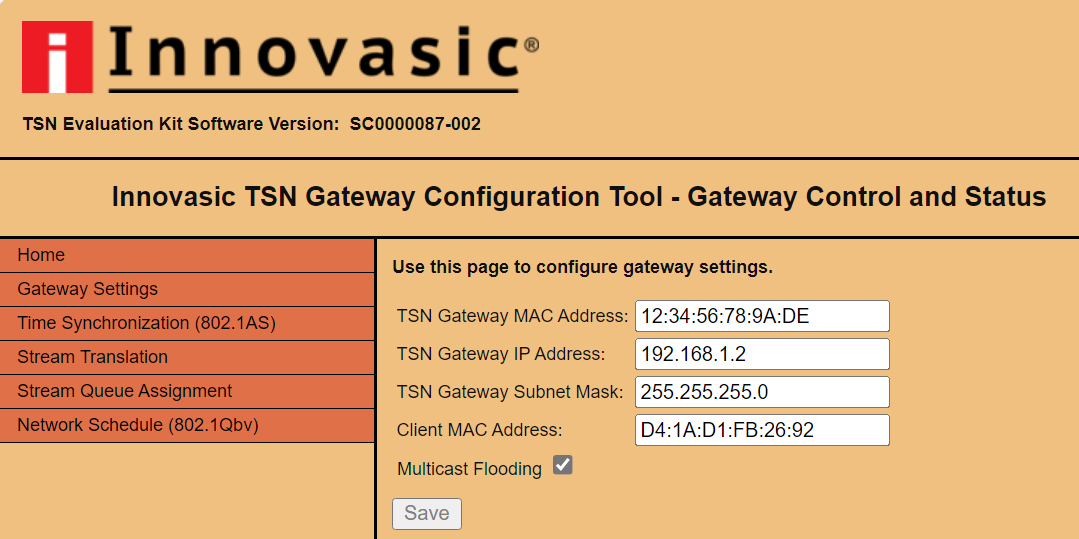}
\caption{TSN gateway GUI for configuration.}
\label{5G_TSN_GUI}
\vspace{-1em}
\end{figure}

The TSN gateway can be configured using a web browser and graphical user interface (GUI) as shown in Fig.~\ref{5G_TSN_GUI}. The web-based GUI is accessible from either of the two TSN ports. The configuration of the TSN gateway as per IEEE 802.1Qbv functionality is described as follow. 

\begin{figure}
\centering
\includegraphics[scale=0.26]{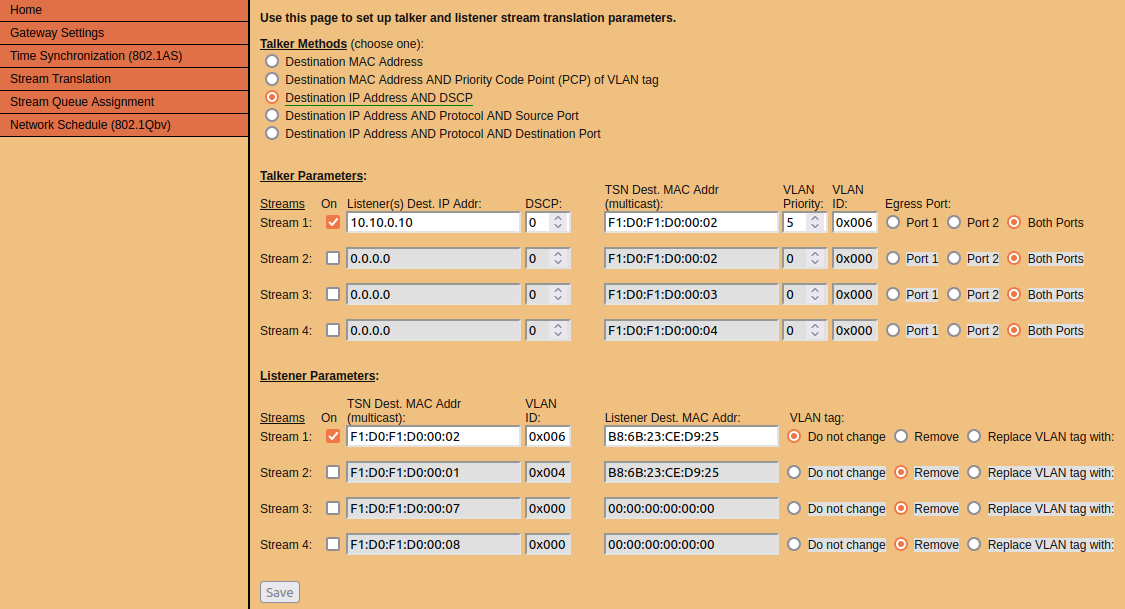}
\caption{Stream translation in the TSN gateway. }
\label{5G_TSN_ST}
\vspace{-1em}
\end{figure}

Once the gateway has been configured with the correct gateway IP and client MAC addresses, we need to identify which streams from the standard Ethernet device need to be input to the TSN network. Streams not identified for use in a TSN network are automatically classified as best-effort traffic and are handled like in any other standard Ethernet network. These streams are not able to use any of the TSN features. Fig.~\ref{5G_TSN_ST} shows the stream configuration for our experiments, where 10.10.0.10 is the IP assigned to the 5G device by the 5G core network.

\begin{figure}
\centering
\includegraphics[scale=0.4]{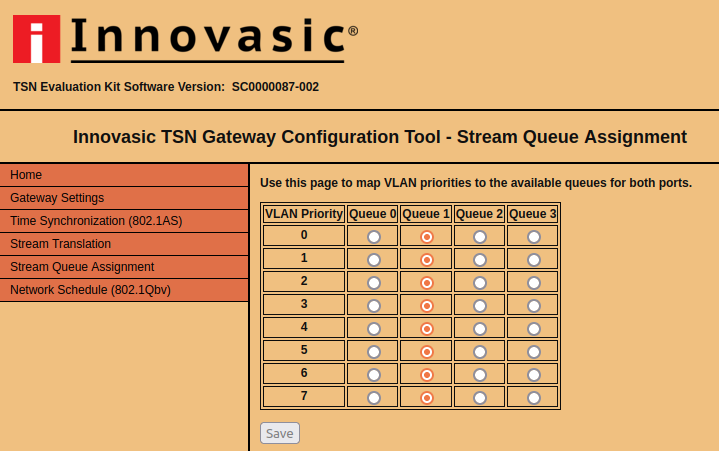}
\caption{TSN gateway stream queue assignment. }
\label{5G_TSN_SQA}
\vspace{-1em}
\end{figure}

Once the streams for the client to send to listeners in the TSN network have been specified, we need to assign these streams to queues for scheduling at specific times on the network. Fig.~\ref{5G_TSN_SQA} illustrates the stream queue assignment process in our experimentation. In our case, all the VLAN priorities have been assigned to Queue 1.

\begin{figure}
\centering
\includegraphics[scale=0.3]{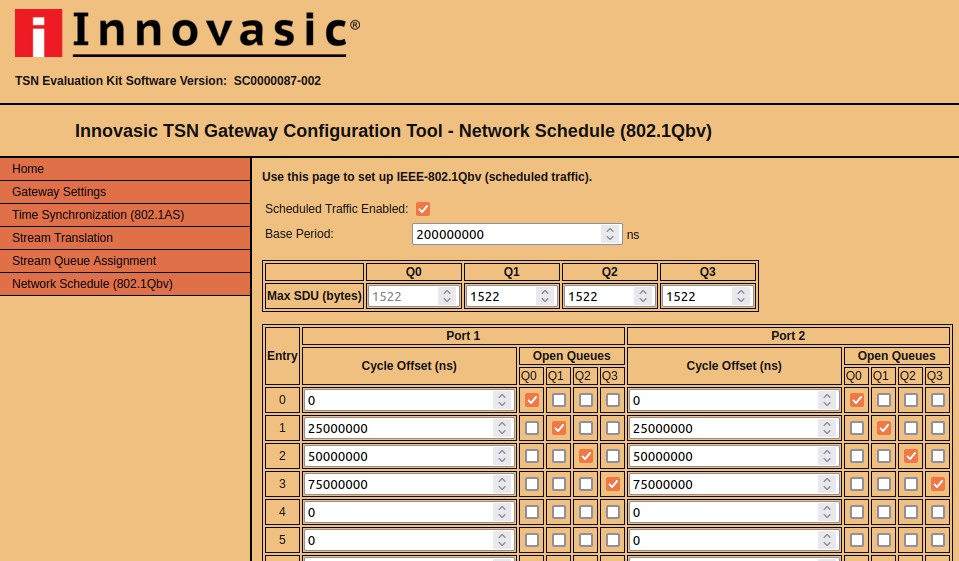}
\caption{TSN gateway queue network schedule. }
\label{5G_TSN_net_sch}
\vspace{-1em}
\end{figure}

After stream queues has been specified, we need to assign these to a schedule. The 802.1Qbv schedule is defined as \emph{transmit windows} within a \emph{base period}. The base period represents the cycle or periodicity after which the schedule repeats. The transmit windows represents the duration of the transmission. The resolution of the schedule is in nanoseconds.  We have used different network schedules for our experiments; one of which is shown in Fig.~\ref{5G_TSN_net_sch}. In this case, we have used the \emph{base period} of 200 msec and \emph{transmit window} of 25 msec. Further, we have assigned queues to the transmit windows.


\begin{figure*}
\centering
\includegraphics[scale=0.45]{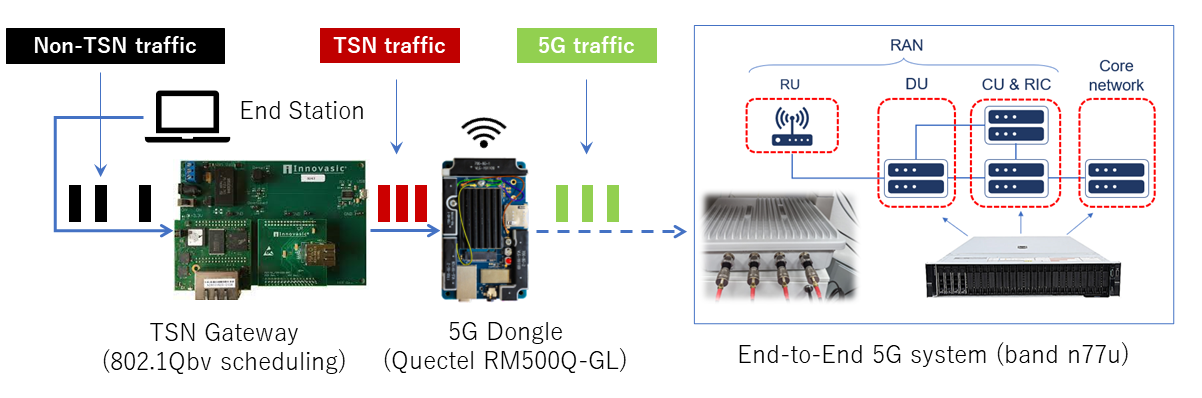} 
\caption{Testbed setup for experimental evaluation. }
\label{testbed}
\vspace{-1em}
\end{figure*}

\section{Experimental Evaluation and Results}\label{sect_exp}
The end-to-end integrated 5G and TSN setup is shown in Fig. \ref{testbed}. Our approach for experimentation is to generate standard Ethernet traffic (non-TSN traffic), translate it to TSN traffic, and transmit it over the air via the 5G dongle to the 5G core network. The non-TSN traffic comes from a laptop (end station) connected to the non-TSN port of the TSN gateway. The 5G dongle is connected to one of the TSN ports of the TSN gateway (via a USB 3.0 to 100 Mbps Ethernet interface). The 5G dongle is also connected (wirelessly) to the 5G network. The non-TSN traffic is generated by iPerf test between the end station and the 5G core network. We investigate the impact of different 802.1Qbv parameters on the end-to-end deterministic performance. We use Wireshark for packet capture at different interfaces. 

As described earlier, through the TSN gateway, we can specify which streams from the standard Ethernet device to translate into the TSN network. These streams are routed from TSN network by tagging the frames with a VLAN tag and assigning a VLAN priority. These streams are then assigned to the queues, and further time windows which dictate transmission as per the schedule.

\begin{figure}[h]
    \centering
    \includegraphics [scale=0.5]{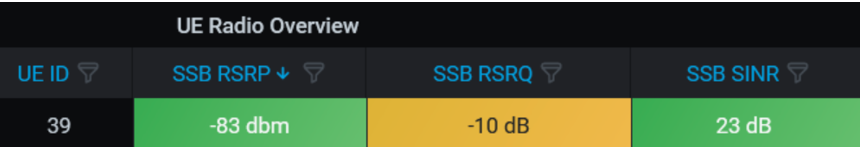}
    \caption{Link-level performance of the 5G dongle.}
    \label{5G_dev_meas}
  \vspace{-1em}
\end{figure}

The 5G system is operating with 40 MHz bandwidth. The DU supports an uplink-centric slot configuration where 5 slots (and 6 symbols) are allocated for the downlink and 4 slots (and 4 symbols) are allocated for the uplink. \emph{For this configuration, the 5G system provides average, maximum, and minimum round-trip latency of 11.36, 33.1, and 6.3 msec \cite{BEACON-5G_CSCN}}. The 5G dongle is successfully connected to the network with link-level performance shown in Fig. \ref{5G_dev_meas}. The link-level performance is sufficiently good (e.g., SINR of 23 dB), highlighting that the transmitted TSN traffic is not significantly affected by impairments on the air-interface. The 5G dongle is provisioned for a specific slice on the 5G system with dedicated user-plane functions in the RAN and the core network.

We define different scenarios with respect to TSN base period and transmit window. An independent iPerf test has been conducted for each scenario.

\begin{figure}
\centering
\includegraphics[width=1\linewidth]{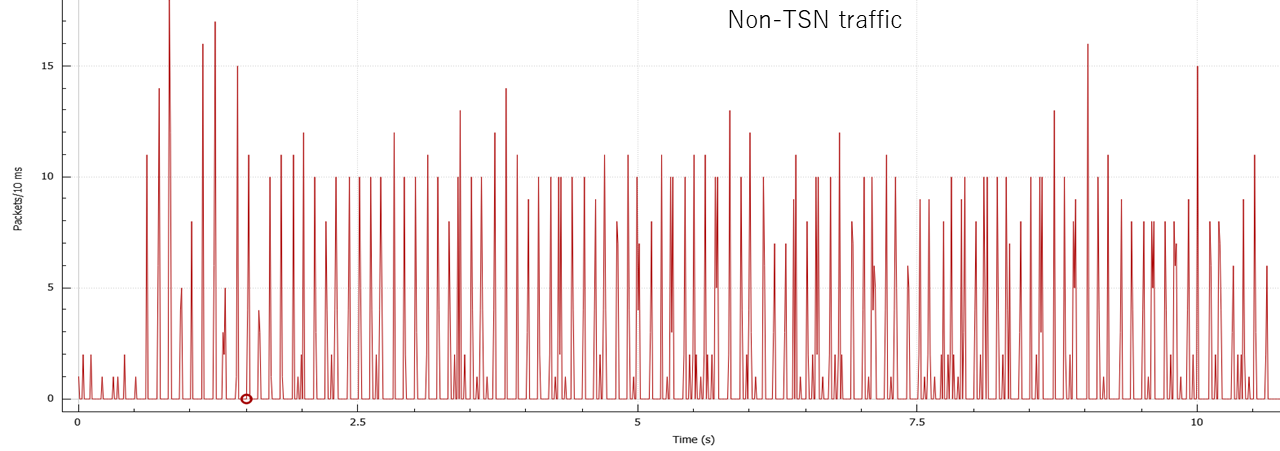}
    \caption{Non-TSN traffic in Scenario 1.}
    \label{non_tsn_200}
    \vspace{-1em}
\end{figure}

\begin{figure}
\centering
\includegraphics[width=1\linewidth]{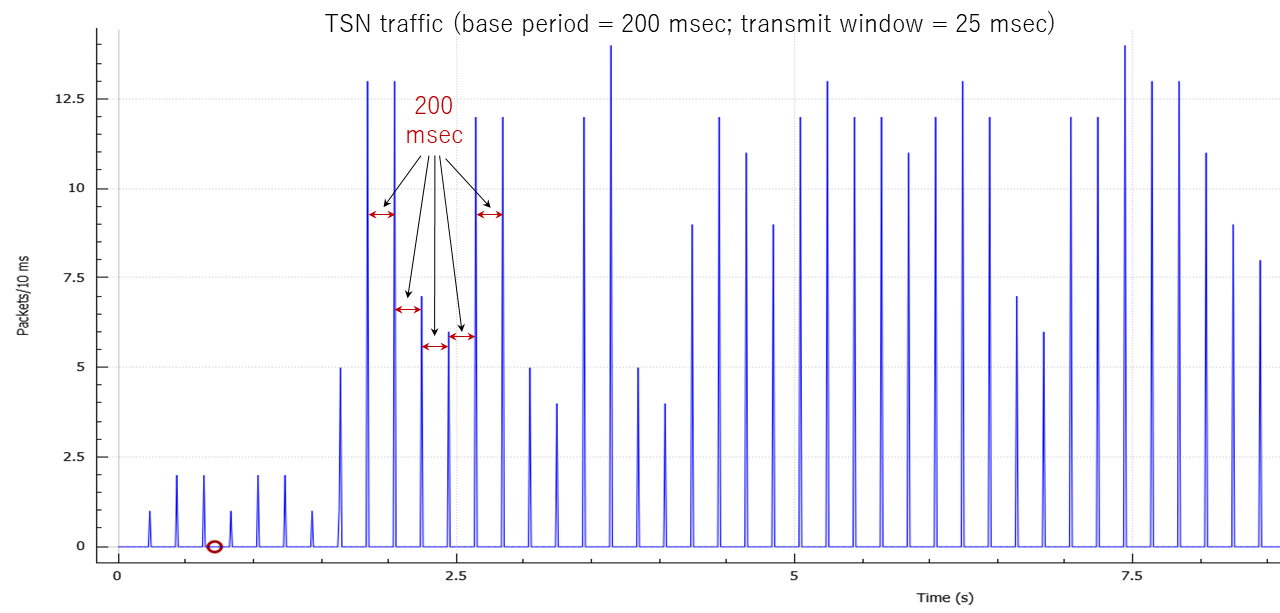}
    \caption{TSN traffic in Scenario 1.}
    \label{tsn_200}
    \vspace{-1em}
\end{figure}

\begin{figure}
\centering
\includegraphics[width=1\linewidth]{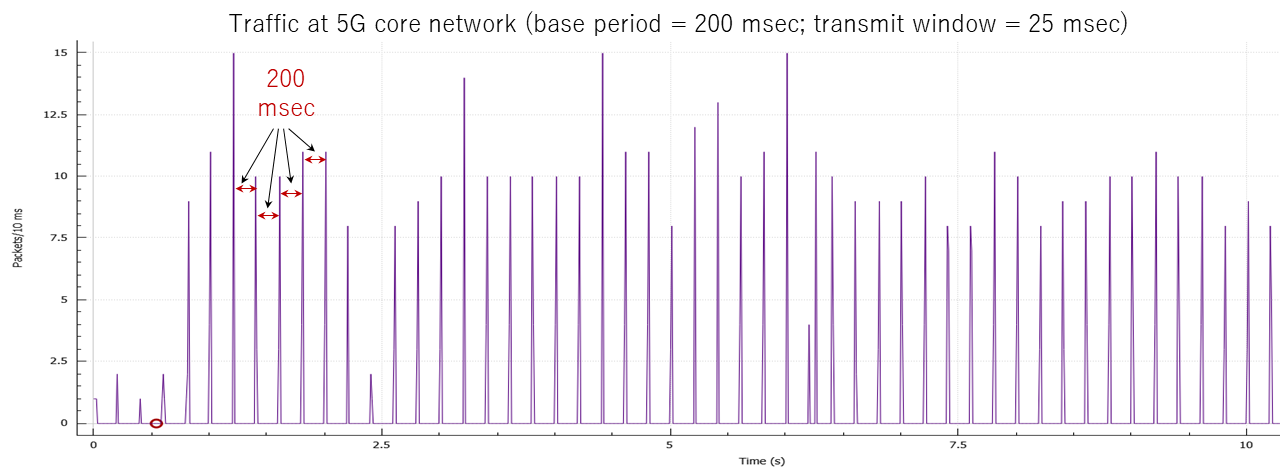}
    \caption{Traffic arriving at the 5G core network in Scenario 1.}
    \label{5G_200}
    \vspace{-1em}
\end{figure}

\textbf{Scenario 1} -- Initially, we set the base period to a relatively large value of 200 msec. The transmit window is set to 25 msec. The non-TSN traffic in this case is shown in Fig. \ref{non_tsn_200}. As this is standard Ethernet traffic, it does not have a regular pattern. Compared to this, the translated TSN traffic, shown in Fig. \ref{tsn_200}, follows strict periodicity of 200 msec. The traffic arriving at the 5G core network, following over-the-air transmission of TSN traffic by the 5G device, is shown in Fig. \ref{5G_200}. This traffic generally shows a periodic pattern with periodicity of 200 msec. This is because the base period is relatively, large compared to the latency of the 5G system, providing sufficient margin for consecutive TSN transmissions. 


\begin{figure}
\centering
\includegraphics[width=1\linewidth]{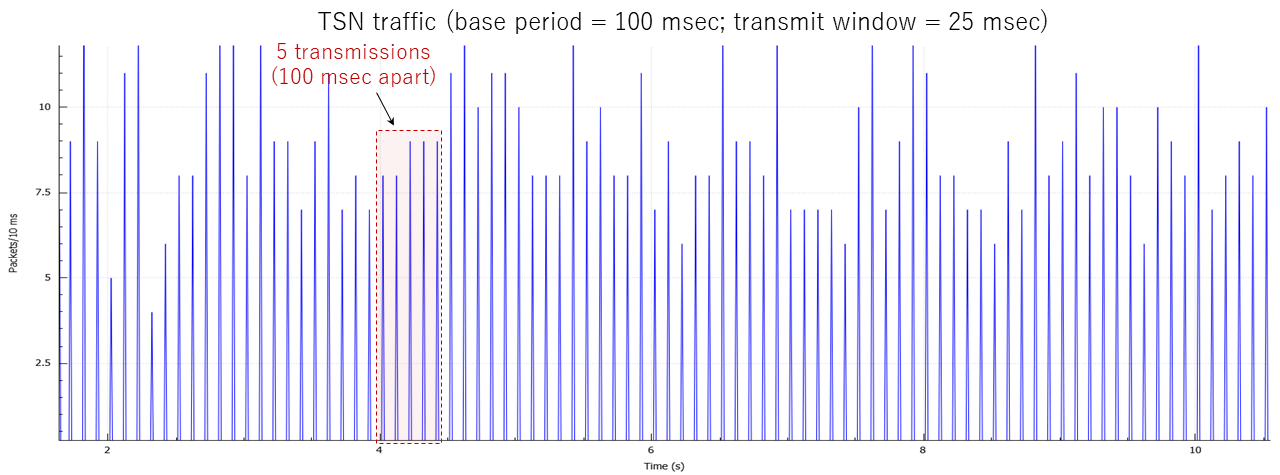}
    \caption{TSN traffic in Scenario 2.}
    \label{tsn_100}
   \vspace{-1em}
\end{figure}

\begin{figure}
\centering
\includegraphics[width=1\linewidth]{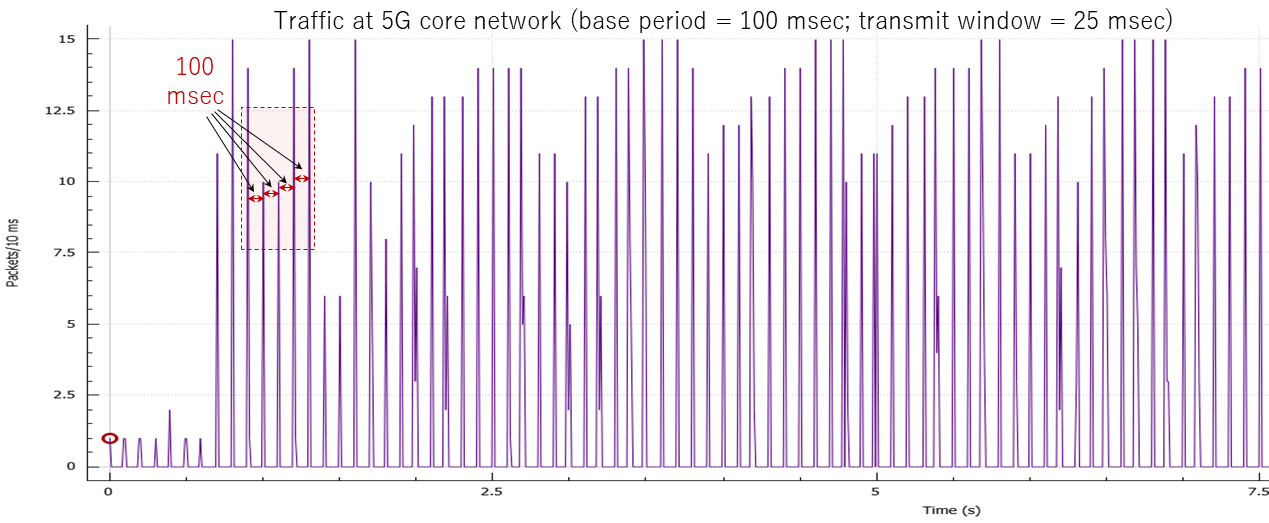}
    \caption{Traffic arriving at the 5G core network in Scenario 2.}
    \label{5G_100}
    \vspace{-1em}
\end{figure}

\textbf{Scenario 2} -- Next, we set the base period to 100 msec and keep the transmit window to 25 msec. The translated TSN traffic and the traffic arriving at the core network in this case is shown in Fig. \ref{tsn_100} and Fig. \ref{5G_100}, respectively. Both show a periodic pattern with periodicity of 100 msec, indicating that the base period is relatively large and there is sufficient margin for the 5G system for handling consecutive TSN transmissions.

\begin{figure}
\centering
\includegraphics[width=1\linewidth]{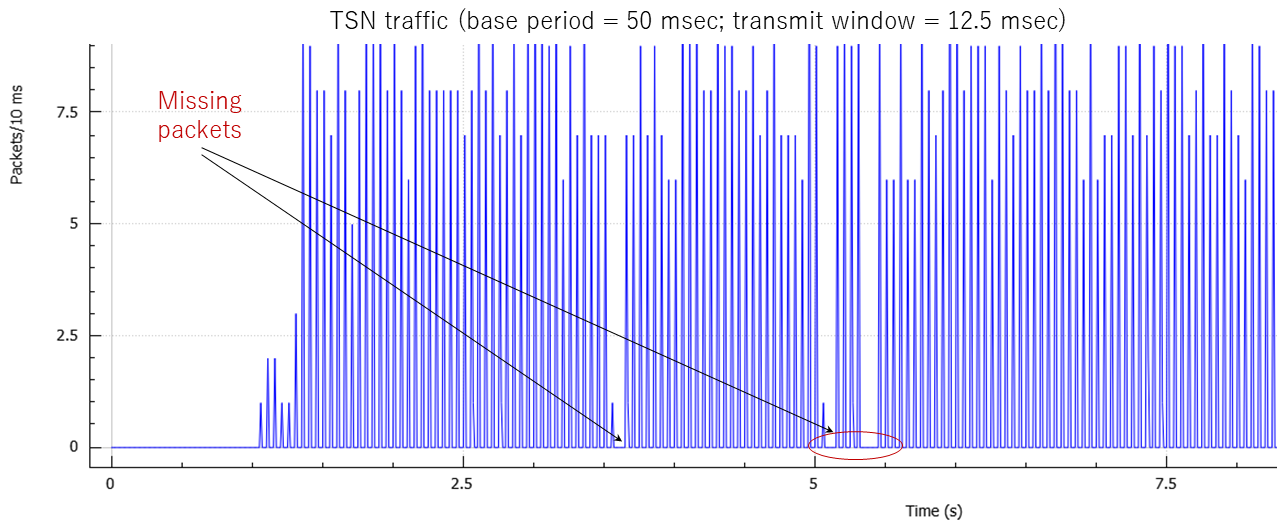}
    \caption{TSN traffic in Scenario 3.}
    \label{tsn_50}
    \vspace{-1em}
\end{figure}

\begin{figure}
\centering
\includegraphics[width=1\linewidth]{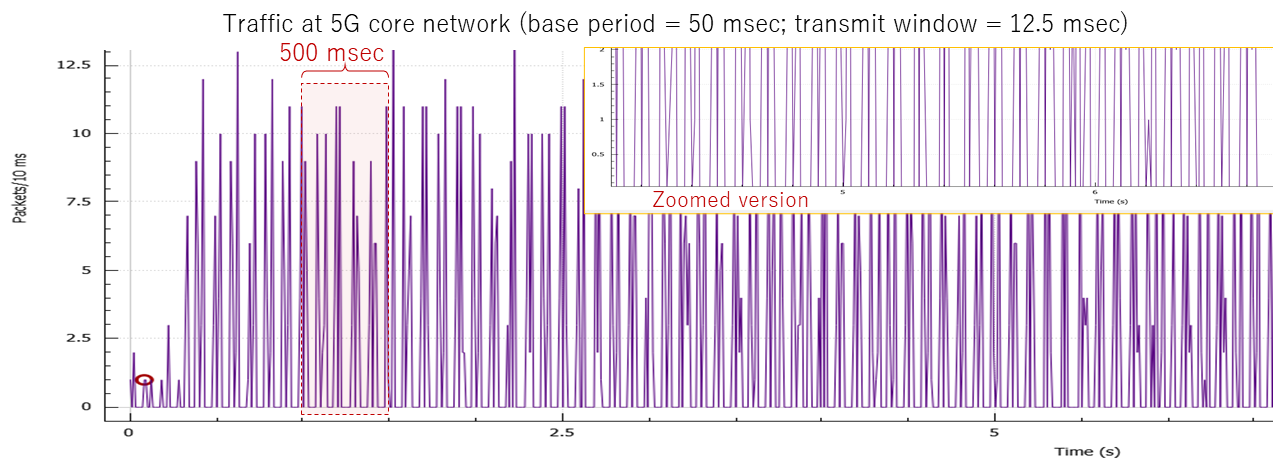}
    \caption{Traffic arriving at the 5G core network in Scenario 3.}
    \label{5G_50}
   \vspace{-1em}
\end{figure}

\textbf{Scenario 3} -- Next, we set the base period to 50 msec and the transmit window to 12.5 msec. Compared to previous scenarios, this scenario creates more stringent traffic requirements. The translated TSN traffic in this case is shown in Fig. \ref{tsn_50}. As expected, it has a periodic behavior with 50 msec periodicity. We observe some missing packets in TSN traffic; however, these are not missing in the corresponding 5G traffic log which implies that these were successfully transmitted but likely not decoded by Wireshark. The traffic arriving at the 5G core network in this case is shown in Fig. \ref{5G_50}. Although the traffic exhibits periodic behavior with 50 msec periodicity, it is not strictly periodic in some instances. The results indicate insufficient margin in the 5G system for effectively handling TSN traffic, in the absence of any TSN-specific optimizations in the RAN or the core network.

\begin{figure}
\centering
\includegraphics[width=1\linewidth]{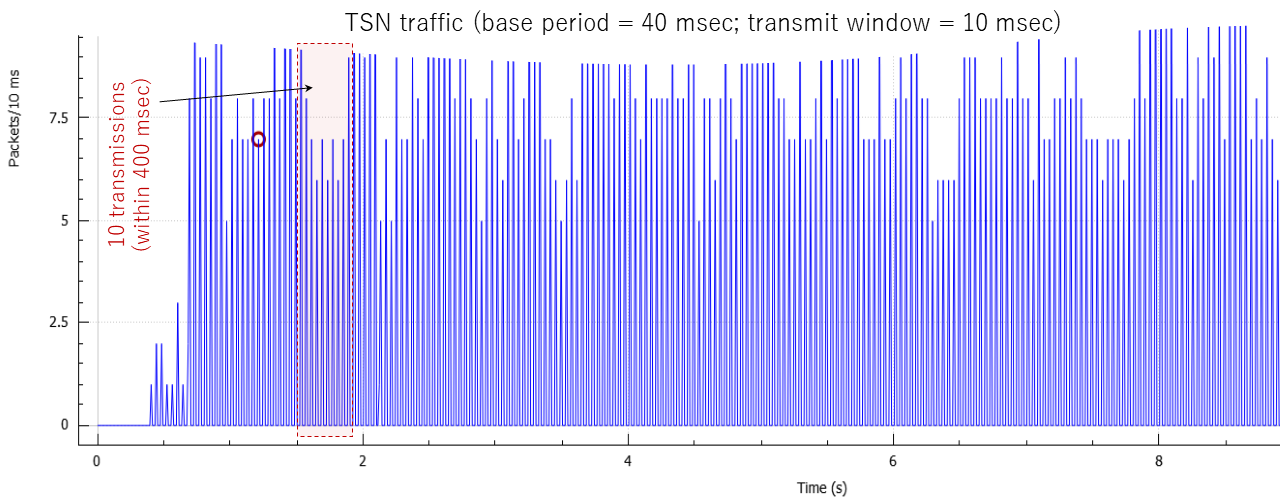}
    \caption{TSN traffic in Scenario 4.}
    \label{tsn_40}
    \vspace{-1em}
\end{figure}

\begin{figure}
\centering
\includegraphics[width=1\linewidth]{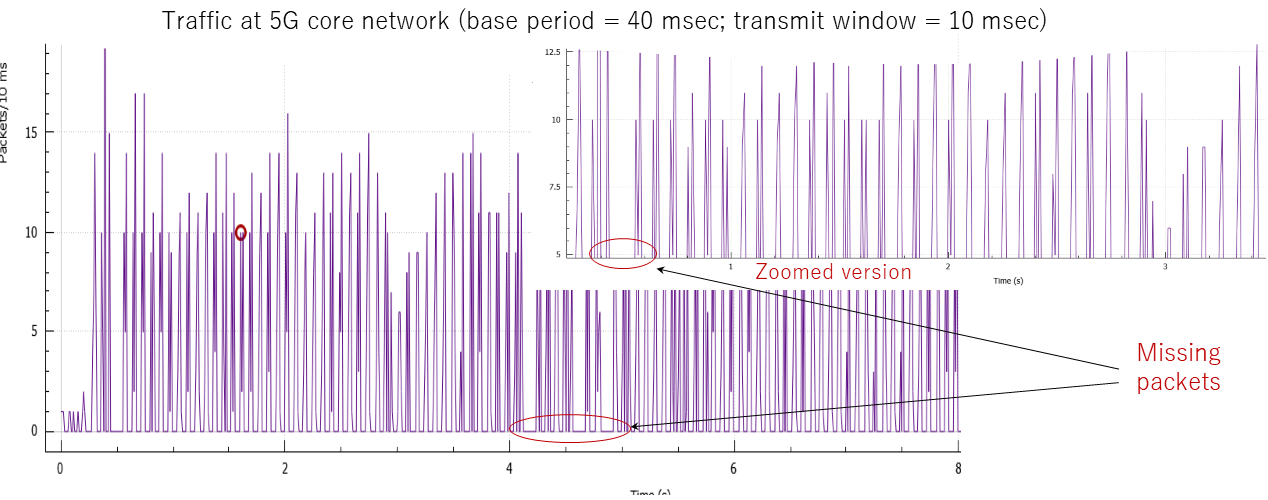}
    \caption{Traffic arriving at the 5G core network in Scenario 4.}
    \label{5G_40}
    \vspace{-1em}
\end{figure}

\textbf{Scenario 4} -- Last but not least, we reduce the base period to 40 msec with a lower transmit window of 10 msec. The translated TSN traffic in this case is shown in Fig. \ref{tsn_40}. It exhibits a periodic behavior with 40 msec periodicity. The traffic arriving at the 5G core network is shown in Fig. \ref{5G_40}. Similar to the previous scenario, the traffic shows a periodic pattern (with 40 msec periodicity); however, it is not strictly periodic in some instances. Additionally, we observe some missing packets which could be either delayed or lost within the 5G system. The results highlight that the vanilla 5G system is not fully equipped to effectively handle TSN traffic with more stringent requirements.

\begin{table}[ht]
\centering
\caption{Parameters for experimental evaluation}
\label{t_parameters}
\begin{tabular}{cc}
\toprule
Parameter             & Value   \\ \midrule
TSN link speed & 100 Mbps \\	
MTU on TSN gateway & 1500 bytes \\
	5G RU configuration & 2 MIMO layers       \\ 
5G RU transmit power & 35 dBm per port \\
EIRP (5G system) & \(-\)5.6 dBm \\
\bottomrule
\end{tabular}
\vspace{-1em}
\end{table}

\section{Key Insights and Takeaways}\label{sect_insights}
The key insights and takeaways from our experimental evaluation are summarized as follows. 

\begin{itemize}
\item Our results indicate that optimal handling of 802.1Qbv traffic for end-to-end deterministic performance in integrated deployments is highly dependent on the native capabilities of the 5G system. A vanilla 5G system is not equipped for handling TSN traffic with stringent requirements. This necessitates end-to-end optimization\footnote{Such optimization techniques have been disclosed in our recent work \cite{patent_virtual_TSN}.} of a 5G system, especially considering the dynamics of TAS.

\item Our evaluation reveals the need for jointly operating an integrated 5G and TSN system as per a global schedule \cite{patent_schedule_system} which takes into account the capabilities of the 5G system, in addition to TSN capabilities and constraints and traffic flow requirements.  

\item Further, the TSN as well as the 5G system must be configured with the global schedule.  In our testbed setup, only the TSN gateway was configured (via the web interface). Such configuration can be done by a central controller residing at the edge of the network. 

\item Tight time synchronization between 5G and TSN systems becomes important for converged operation as per a global schedule. Our hybrid 5G-TSN system is loosely time synchronized. Nevertheless, the impact of time synchronization on end-to-end deterministic performance needs to be evaluated. 

\item Our testbed is based on off-the-shelf TSN and 5G devices. Although these were successfully integrated for over-the-air transmission of TSN traffic, there is a risk for accumulating uncontrolled software delays affecting end-to-end deterministic performance. Devices with built-in TSN and 5G capabilities are important for mitigating this risk.

\item The DU in our 5G system does not implement TSN-aware packet scheduling techniques. In the absence of such techniques, providing end-to-end deterministic performance becomes challenging especially when the TSN traffic requirements become more stringent. Window-based scheduling in 5G \cite{patent_schedule} is a promising technique for effectively handling TSN traffic in a deterministic manner. 

\item The programmability and flexibility offered by the O-RAN architecture \cite{o2021ran} is promising for optimizing the RAN with protocols for TSN-like functionality as well as for joint configuration and management of integrated 5G and TSN deployments. 

\end{itemize}

\section{Conclusion}\label{sect_cr}
Convergence of 5G and TSN systems is an important step toward adoption of standardized technologies in industrial networks. This paper conducted a testbed-based evaluation of a hybrid 5G and TSN system with 802.1Qbv traffic for end-to-end deterministic performance. The evaluation shows successful TSN traffic translation and its over-the-air transmission via the 5G dongle. In terms of performance, deterministic traffic patterns are observed in 5G traffic when the base period of TSN traffic is relatively higher than the average latency offered by the 5G system. However, pseudo-deterministic behaviour is observed with smaller base periods comparable to the average latency performance of the 5G system. The results further reveal that a vanilla 5G system may not be equipped to effectively handle TSN traffic, especially with stringent requirements. The shortcomings can manifest in the form of packet losses or delays. Hence, there is a need for end-to-end optimization of a 5G system as per TSN traffic dynamics and protocols that provide TSN-like functionality while the 5G system acts as a black box to TSN entities.

%

\bibliographystyle{IEEEtran}
\bibliography{bibliography.bib}
\end{document}